\documentclass[twocolumn,reprint, superscriptaddress,amsmath,amssymb,aps,]{revtex4-2}
\usepackage[utf8]{inputenc}
\usepackage{graphicx}
\usepackage{dcolumn}
 
\usepackage{bm}
\usepackage{dirtytalk}
\usepackage{siunitx}
\usepackage{amssymb}
\usepackage{standalone}

\begin{document}


\title{Imaging of electrically controlled van der Waals layer stacking in 1$T$-TaS$_2$}

\author{Corinna Burri}
\affiliation{PSI Center for Photon Science, Paul Scherrer Institute, 5232 Villigen PSI, Switzerland}
\affiliation{Laboratory for Solid State Physics and Quantum Center, ETH Zurich, 8093 Zurich, Switzerland}

\author{Nelson Hua}
\author{Dario Ferreira Sanchez} 
\affiliation{PSI Center for Photon Science, Paul Scherrer Institute, 5232 Villigen PSI, Switzerland}

\author{Wenxiang Hu}
\author{Henry~G.~Bell}
\affiliation{PSI Center for Photon Science, Paul Scherrer Institute, 5232 Villigen PSI, Switzerland}
\affiliation{Laboratory for Solid State Physics and Quantum Center, ETH Zurich, 8093 Zurich, Switzerland}

\author{Rok~Venturini}
\affiliation{PSI Center for Photon Science, Paul Scherrer Institute, 5232 Villigen PSI, Switzerland}
\affiliation{Department of Complex Matter, Jozef Stefan Institute, 1000 Ljubljana, Slovenia}

\author{Shih-Wen~Huang} 
\affiliation{PSI Center for Photon Science, Paul Scherrer Institute, 5232 Villigen PSI, Switzerland}

\author{Aidan G. McConnell}
\author{Faris~Dizdarević}
\affiliation{PSI Center for Photon Science, Paul Scherrer Institute, 5232 Villigen PSI, Switzerland}
\affiliation{Laboratory for Solid State Physics and Quantum Center, ETH Zurich, 8093 Zurich, Switzerland}

\author{Anže~Mraz}
\affiliation{Department of Complex Matter, Jozef Stefan Institute, 1000 Ljubljana, Slovenia}
\affiliation{CENN Nanocenter, 1000 Ljubljana, Slovenia}
\author{Damjan~Svetin}
\affiliation{Department of Complex Matter, Jozef Stefan Institute, 1000 Ljubljana, Slovenia}

\author{Benjamin Lipovšek}
\author{Marko Topič}
\affiliation{Faculty for Electrical Engineering, University of Ljubljana, 1000 Ljubljana, Slovenia}

\author{Dimitrios~Kazazis} 
\affiliation{PSI Center for Photon Science, Paul Scherrer Institute, 5232 Villigen PSI, Switzerland}

\author{Gabriel~Aeppli}
\affiliation{PSI Center for Photon Science, Paul Scherrer Institute, 5232 Villigen PSI, Switzerland}
\affiliation{Laboratory for Solid State Physics and Quantum Center, ETH Zurich, 8093 Zurich, Switzerland}
\affiliation{Institute of Physics, EPF Lausanne, 1015 Lausanne, Switzerland}  

\author{Daniel Grolimund}
\affiliation{PSI Center for Photon Science, Paul Scherrer Institute, 5232 Villigen PSI, Switzerland}

\author{Yasin Ekinci} 
\affiliation{PSI Center for Photon Science, Paul Scherrer Institute, 5232 Villigen PSI, Switzerland}

\author{Dragan Mihailovic}
\email{dragan.mihailovic@ijs.si}
\affiliation{Department of Complex Matter, Jozef Stefan Institute, 1000 Ljubljana, Slovenia}
\affiliation{CENN Nanocenter, 1000 Ljubljana, Slovenia}
\affiliation{Faculty of Mathematics and Physics, University of Ljubljana, 1000 Ljubljana, Slovenia}

\author{Simon Gerber}
\email{simon.gerber@psi.ch}
\affiliation{PSI Center for Photon Science, Paul Scherrer Institute, 5232 Villigen PSI, Switzerland}

\date{\today}

\begin{abstract}
{\textbf{Van der Waals (vdW) materials exhibit a variety of states that can be switched with low power at low temperatures, offering a viable cryogenic “flash memory” required for the classical control electronics for solid-state quantum information processing. In 1$T$-TaS$_2$, a non-volatile metallic ‘hidden’ state can be induced from an insulating equilibrium charge-density wave ground state using either optical or electrical pulses. Given that conventional memristors form localized, filamentary channels which support the current, a key question for design concerns the geometry of the conduction region in highly energy-efficient 1$T$-TaS$_2$ devices. Here, we report \textit{in operando} micro-beam X-ray diffraction, fluorescence, and concurrent transport measurements, allowing us to spatially image the non-thermal hidden state induced by electrical switching of \mbox{1$T$-TaS$_2$}. Our results reveal a long-range ordered, non-filamentary switched state that extends well below the electrodes, implying that the self-organized, collective growth of the hidden phase is driven by a combination of charge flow and lattice strain. Our unique combination of techniques showcases the potential of non-destructive, three-dimensional X-ray imaging to study bulk switching properties in microscopic detail, namely electrical control of the vdW layer stacking.}
}
\end{abstract}
\maketitle

Switching between low and high resistance states via electrical excitation has significant potential for analog and neuromorphic computing applications~\cite{Rao1989,Boahen2006,Yang2013}. A key requirement for this is the ability to reversibly transition between two resistance states which can occur through mechanisms such as the formation of a conductive filament, ferroelectric and magnetic tunneling junctions, or phase transitions between crystalline and amorphous states \cite{Yang2013, Garcia2014,Nilson2020, Gallo2020}. There is substantial research primarily focused on identifying new candidates for fast and energy-efficient memory devices, and also cryo-computing~\cite{Holmes2013}. Van der Waals~(vdW) materials, due to their layered structure, are well-suited not just to explore novel states and phase \mbox{transitions~\cite{Novoselov2000,Wang2015,Cao2018,Kennes2021}} but are potentially also useful for scalable electronic devices~\cite{Wang2012,Jariwala2014,Liu2016}. 

Among vdW materials, the transition metal dichalcogenide 1$T$-TaS$_2$ attracts considerable attention for its unique combination of correlated electron phenomena. These include various charge-density wave (CDW) states~\cite{Wilson1975,Fazekas1979}, superconductivity under pressure or doping~\cite{Sipos2008,Li2012}, and a putative quantum spin liquid \mbox{phase~\cite{Klanjsek2017,Law2017}}. At low temperatures it is believed to exhibit a Mott insulating state~\cite{TOSATTI1976,Sipos2008}, characterized by commensurate~(C) CDW order formed by star-shaped polaron domains that tessellate the layers below \mbox{$150-180$~K~\cite{Scruby1975,TOSATTI1976}}. Moreover, and of greatest interest to us here, it shows non-thermal, reversible switching to a metastable, metallic `hidden'~(H) CDW phase upon application of ultrashort optical or electrical pulses. The lifetime of this HCDW state is short at elevated temperatures and stable below $\approx40$~K~\cite{Stojchevska,Vaskivskyi20152,Hollander2015,Yoshida2015,Vaskivskyi2016}.

The microscopic origin of the non-thermal switching of  1$T$-TaS$_2$ has been intensely studied but is not fully understood~\cite{Stojchevska,Cho2016,Ma2016,Vaskivskyi20152,Gerasimenko2019,Stahl2020,Mihailovic2021,Torre2021}. It can be described in terms of a low-temperature free-energy landscape that features a global minimum (the Mott insulating equilibrium state) and multiple local minima separated by potential energy barriers~\cite{,Stojchevska,Vaskivskyi20152,Torre2021, Mihailovic2021}. Excitation with a laser or current pulse can drive the system into these metastable states.

Optical switching of 1$T$-TaS$_2$ with femtosecond laser pulses, exciting electron and holes homogeneously, has been extensively studied with both \mbox{quasi-static \cite{Gerasimenko2019, Gerasimenko20192,Vodeb2019}} and time-resolved~\mbox{\cite{Perfetti2006,Stojchevska,Ravnik2018,Zong2018,Stahl2020,Danz2021,Maklar2022,Liu2024,Hua2024}} techniques. Surface-sensitive scanning tunneling microscopy~(STM) shows formation of conducting domain walls~\cite{Stojchevska, Gerasimenko2019}, while bulk-sensitive X-ray diffraction (XRD) reveals changes in the out-of-plane polaron stacking and a different in-plane commensurability of the HCDW~state compared to the low- and room-temperature CDW states, confirming a non-thermal switching mechanism~\cite{Stahl2020,Hua2024}.

In contrast, electrical switching has been less explored despite its promise for application in highly-efficient cryo-memory devices due to the compatibility with standard electronics. The electrically-induced hidden \mbox{(e-HCDW)} state has been studied using transport measurements~\mbox{\cite{Yoshida2015,Hollander2015,Mraz2022,Venturini2022}}, STM~\cite{Vaskivskyi2016,Ma2016,Cho2016}, and angle-resolved photoemission spectroscopy~\cite{Nitzav2024}, revealing macroscopic behavior similar to the optically-induced hidden (o-HCDW) state. Domain wall formation at the surface has also been observed with STM, but unlike the o-HCDW, it is understood that the e-HCDW state results from excitation due to charge carrier separation, leading to more inhomogeneous domain walls~\cite{Vaskivskyi2016,Mihailovic2021}. Furthermore, in contrast to the \mbox{o-HCDW} state, the \mbox{e-HCDW} can be triggered with much longer pulses up to \SI{100}{\milli\second}~\cite{Svetin2017}, suggesting key differences in their switching mechanisms. As previously discussed~\cite{Vaskivskyi2016,Mihailovic2021}, the switching is non-thermal even for long current pulses.

It is unclear whether optical and electrical excitations lead to the same local free-energy minimum. If they do, can these two distinct perturbations---differing in pulse duration, electron-hole pairs accessed, and directionality---follow the same non-thermal pathway, or are there different mechanisms at play?
Also, the lack of spatially-resolved studies on the e-HCDW leaves the question unanswered whether the switching is surface-localized or propagates through the bulk. Likewise, it is unknown if the \mbox{e-HCDW} spreads across the entire region in-between the electrodes or forms conducting filament channels. Finally, it is not established whether the principal actor of the switching is the electric field or the current. Resolving the spatial structure and population of the e-HCDW would provide insights into the switching mechanisms, and thereby contribute to designing devices with improved performance and scalability.

The required information can be obtained by mapping the bulk CDW states before and after switching. If the e-HCDW is structurally equivalent to the o-HCDW, then the known CDW diffraction peaks from photo-driven experiments~\cite{Stahl2020} can be used as a fingerprint to identify the e-HCDW switching region. Here we report on  \textit{in situ} resistance measurements with micro-beam X-ray fluorescence~($\mu$XRF) and diffraction~($\mu$XRD) to spatially image a nano-fabricated 1$T$-TaS$_2$ device \textit{in operando} (Fig.~\ref{fig_1}). Analysis of CDW Bragg peaks reveals the detailed three-dimensional~(3D) spatial evolution of the material properties before and after electrical excitation.

\section*{Results}

\begin{figure*}[h]
  \centering
    \includegraphics[width=\linewidth, trim=0cm 0cm 0cm 0cm, clip=true]{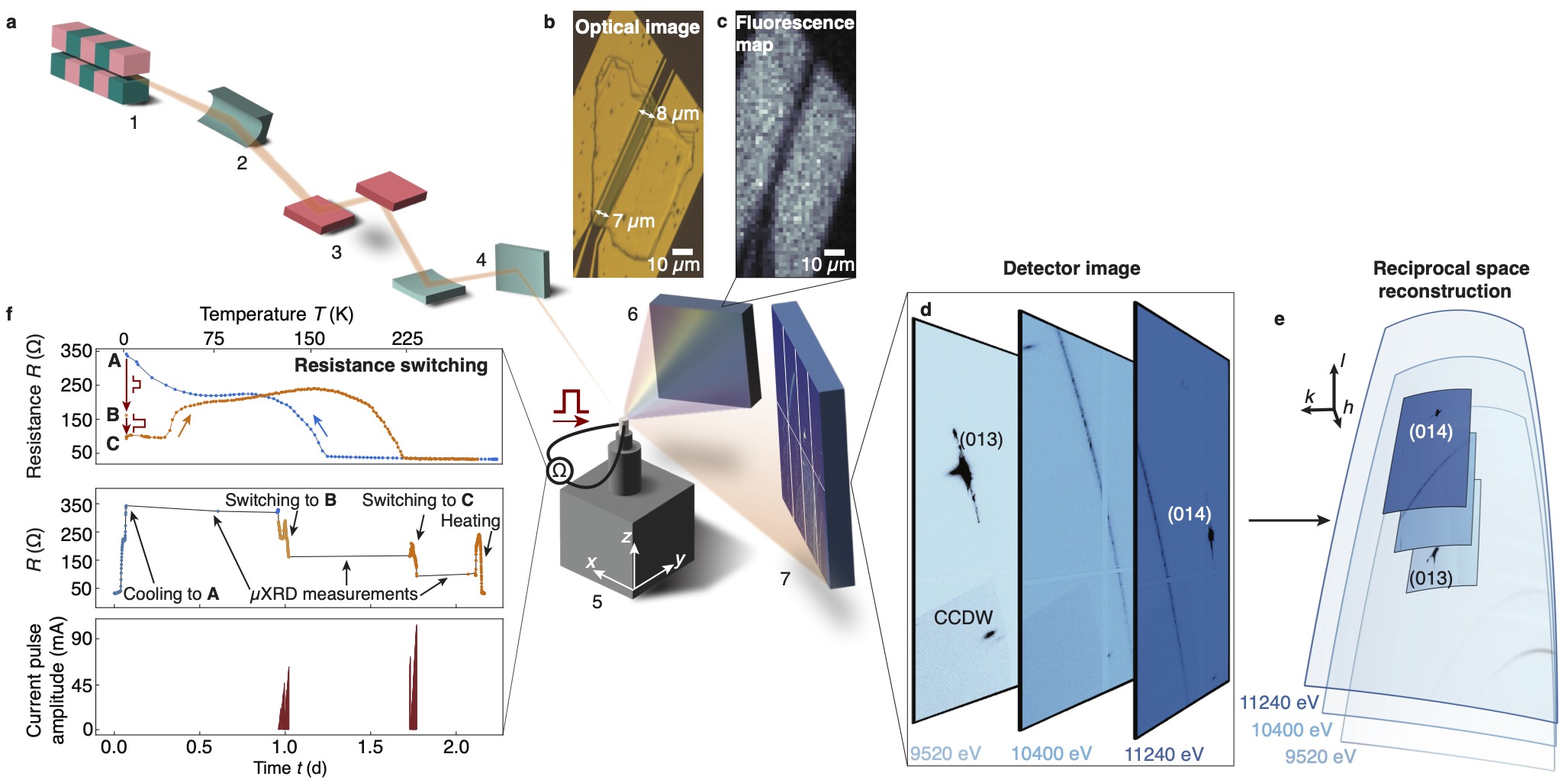}
    \caption{\textbf{\textit{In operando} macro- and microscopic measurement of phase switching}. 
    \textbf{a} Schematic of the synchrotron beamline including the undulator (1), a toroidal mirror (2), the monochromator (3) and Kirkpatrick-Baez focusing mirrors~(4). The 1.5 $\times$ \SI{2.5}{\micro\meter\squared} sized X-ray beam is directed at the 1$T$-TaS$_2$ device (5) in a $^4$He cryostat that can be moved using translation stages. The device is electrically contacted, allowing for resistance measurements and application of current pulses. X-ray fluorescence (6) and diffraction (7) is recorded simultaneously on respective detectors. \textbf{b} Optical image of the \mbox{1$T$-TaS$_2$}~device. \textbf{c}  Au fluorescence map highlighting the electrodes. \textbf{d} Diffraction patterns measured at \SI{6}{\kelvin} at selected X-ray energies with the lattice (013) and (014) Bragg reflections, as well as one CCDW peak. \textbf{e} Conversion of the 2D~detector images taken at various X-ray energies to 3D (\textit{h, k, l}) reciprocal space. \textbf{f} \textit{In situ} resistance as a function of temperature and time with \textbf{A} the unswitched CCDW state, \textbf{B} and \textbf{C} the partially- and fully-switched HCDW~states, respectively.}
    \label{fig_1}
\end{figure*}

Figure~\ref{fig_1}b shows an optical image of the 1$T$-TaS$_2$ device, formed by a contacted flake, used to investigate the e-HCDW state (fabrication details in Methods). The distance between the electrodes was set to~$\approx8~\mu$m, considering the $\mu$m-sized X-ray spot and that the switching voltage scales with the distance between the electrodes~\cite{Mraz2022}. There are two inert contacts between the outer electrodes that are not used for the resistance measurement.
The device is located using $\mu$XRF (Fig.~\ref{fig_1}c) which is sensitive to even small traces of embedded elements~\cite{DAnna2023,Masteghin2024} (Supplementary Information). Before and after applying current pulses, $\mu$XRD and $\mu$XRF measurements are taken at room temperature and \SI{6}{\kelvin}, as well as different energies spanning an extended portion of reciprocal space (Fig.~\ref{fig_1}d,e). Conversion to 3D reciprocal space allows us to identify the equilibrium and newly appearing signals, as well as to determine their position and peak shapes  (Supplementary Information).
Temperature-dependent resistance of the device is measured in a two-probe configuration (Fig.~\ref{fig_1}f). Upon cooling, the flake goes through a first-order phase transition, resulting in a step-like increase of the resistance at $\approx$\SI{150}{\kelvin}. We denote the unswitched low-temperature CCDW state with \textbf{A}. To induce the \mbox{e-HCDW} state, single square current pulses with a width of \SI{100}{\micro\second} are applied. The pulse amplitude is increased gradually until a resistance drop occurs. We denote the state \textbf{B} with $47 \%$ resistance compared to \textbf{A} as partially-switched, reached by applying a pulse amplitude of~\SI{63}{\milli\ampere}. Increasing the pulse height to~\SI{105}{\milli\ampere} induces another resistance drop to $27 \%$ of \textbf{A}, which we refer to as fully-switched~\textbf{C}. Following the resistance upon heating, we observe the characteristic relaxation from the HCDW state to the equilibrium high-temperature states, as well as the first-order, hysteretic transition~\cite{Vaskivskyi20152}, confirming that the HCDW state is induced in the device. As a function of time, the device resistance exhibits changes upon cooling, heating, and current pulse application, else it is stable over hours. During switching, we observe some resistance increase, likely due to a partial relaxation to the CCDW state, and then, as the current amplitude is further increased, a resistance drop indicating that the HCDW state is induced. We note that devices with an optimized design can be switched with a single current pulse \cite{Mraz2022, Venturini2022}, whereas with our device design, optimized for \textit{in situ} $\mu$XRD measurements, multiple pulses with increasing amplitude have to be applied. During the $\mu$XRD measurements at a constant temperature of \SI{6}{\kelvin} the resistance remains stable. Importantly, the resistances measured at room temperature before and after the switching and $\mu$XRD measurements are identical, which directly proves the non-destructive nature of our technique.

Several structural changes occur in the non-equilibrium o-HCDW state compared to the equilibrium CCDW order: (i) the CCDW and `dimer' peaks vanish~\cite{Stahl2020} (the latter are associated with inter-layer dimerization of the star-like domains), (ii) a new long-range order (at different positions in reciprocal space) appears as HCDW peaks~\cite{Stahl2020,Gerasimenko2019}, and (iii) the out-of-plane lattice constant contracts~\cite{Vaskivskyi2024} (observed as a shift of the lattice peaks in the out-of-plane direction). In the following, we show that these three characteristic features are also observed for the e-HCDW state. Here, we focus on the HCDW signal, whereas the vanishing dimer CCDW peaks and the out-of-plane lattice contraction are addressed in the Supplementary Information.  

\begin{figure*}[h]
  \centering
  \includegraphics[width=0.5\linewidth, trim=0cm 0cm 0cm 0cm, clip=true]{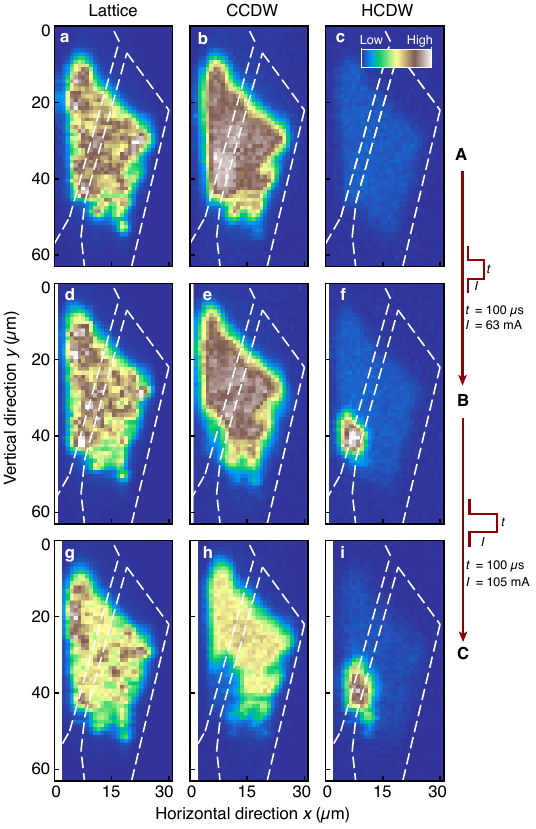}
  \caption{\textbf{Electrically-induced non-thermal phase switching}.
  \textbf{a-c} Spatially-resolved intensity at the reciprocal space positions of the (013) lattice reflection, as well as a nearby CCDW and HCDW peak in the unswitched state \textbf{A}. The lattice and CCDW peaks are observed across the whole flake, whereas no HCDW signal is found. \textbf{d-f} and \textbf{g-i} show the respective peak intensities in the partially- \textbf{B} and fully-switched state \textbf{C}. The CCDW signal is suppressed in the lower left corner of the flake where the HCDW signal emerges. Color scales show a minimal intensity of 0 counts and maxima of~$2.5 \cdot 10^3$~(lattice), $1.3 \cdot 10^3$~(CCDW) and $8.0 \cdot 10^2$ (HCDW) for an integration time of \SI{100}{\milli\second}. Dashed lines indicate the position of the electrodes. }
  \label{fig_2}
\end{figure*}

For each position on the device, we reconstruct the 3D reciprocal space and look for a decrease in intensity of the CCDW peak, as well as newly appearing signals hinting at the e-HCDW state: three 3D regions of interest (ROI) around the (013) lattice, as well as nearby CCDW and o-HCDW positions~\cite{Stahl2020} are set, and the intensity within these ROIs is integrated. The shape of the flake can be clearly seen in the lattice peak maps (Fig.~\ref{fig_2}a,d,g), which remains intact throughout the switching process. However, a change occurs in the CDW structure of the flake: the CCDW peak map before switching (\textbf{A}, Fig.~\ref{fig_2}b) shows the same outline as the lattice peak; but Fig.~\ref{fig_2}e,h in the \mbox{partially- and fully-switched states, \textbf{B}} and \textbf{C}, respectively, reveal that the CCDW peak intensity in the bottom left part of the flake vanishes. Concurrently, an e-HCDW peak signal appears in the same region in-between the electrodes (Fig.~\ref{fig_2}f,i). That is, not the entire flake switches from the CCDW to the HCDW order, but rather a conducting channel appears between the electrodes on one edge of the flake, while the rest of the device remains in the equilibrium state.

\begin{figure*}[h]
  \centering
      \includegraphics[width=0.75\linewidth, clip=true]{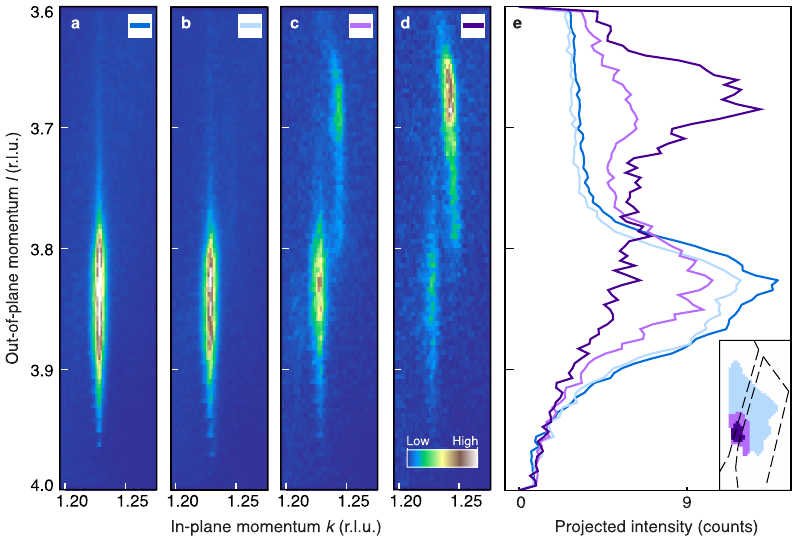}
     \caption{\textbf{Momentum- and real-space structure of the CDW states}. 
     \textbf{a} 2D (\textit{k, l}) projection in reciprocal lattice units~(r.l.u.) of the CCDW~peak in the unswitched state \textbf{A} (dark blue), obtained by integrating the intensity along the in-plane \textit{h} direction and normalizing per pixels. \textbf{b} Measurement of the fully-switched state \textbf{C} in the light blue region (inset of~e), showing a CCDW and only a very faint HCDW signal. \textbf{c,d} Respective projection measured in the light and dark purple regions (inset of~e), close to and in the vicinity of the electrode gap, respectively. Fewer pixels in those regions result in poorer statistics of the projections. \textbf{e} Out-of-plane projection of the data shown in a-d. Dashed lines on the spatial map in the inset indicate the location of the electrodes.}
    \label{fig_3}
\end{figure*} 

Having identified the spatial region of the flake that switches to the HCDW state, we examine the momentum-space structure and real-space distribution of the CDW states. The CCDW signal in Fig.~\ref{fig_3}a features the characteristic elongation parallel to the out-of-plane~$l$ direction due to partial disorder, and the reciprocal space position is close to the previously reported values \mbox{\cite{Ishiguro1991,Lauhle2015,Stahl2020}} (Tab.~\ref{table:Table1}).
\mbox{Figures~\ref{fig_3}b-d} are from the fully-switched state~\textbf{C}, deduced from different regions on the flake and revealing that CCDW order gradually gives way to the e-HCDW state as the switching channel is approached. The reciprocal space position of the \mbox{e-HCDW} peak in-between the electrodes is also found to be consistent with the o-HCDW signal in the literature~\cite{Stahl2020} (Tab.~\ref{table:Table1}). Therefore, we conclude that the optically- and electrically-driven HCDW order are equivalent not only from an electronic but also a 3D structural point of view. In turn, this also implies control of the switching into the same local minimum of the free-energy landscape regardless of the excitation method.


\begin{figure*}[h]
  \centering
      \includegraphics[width=\linewidth,trim=0cm 0cm 0cm 0cm, clip=true]{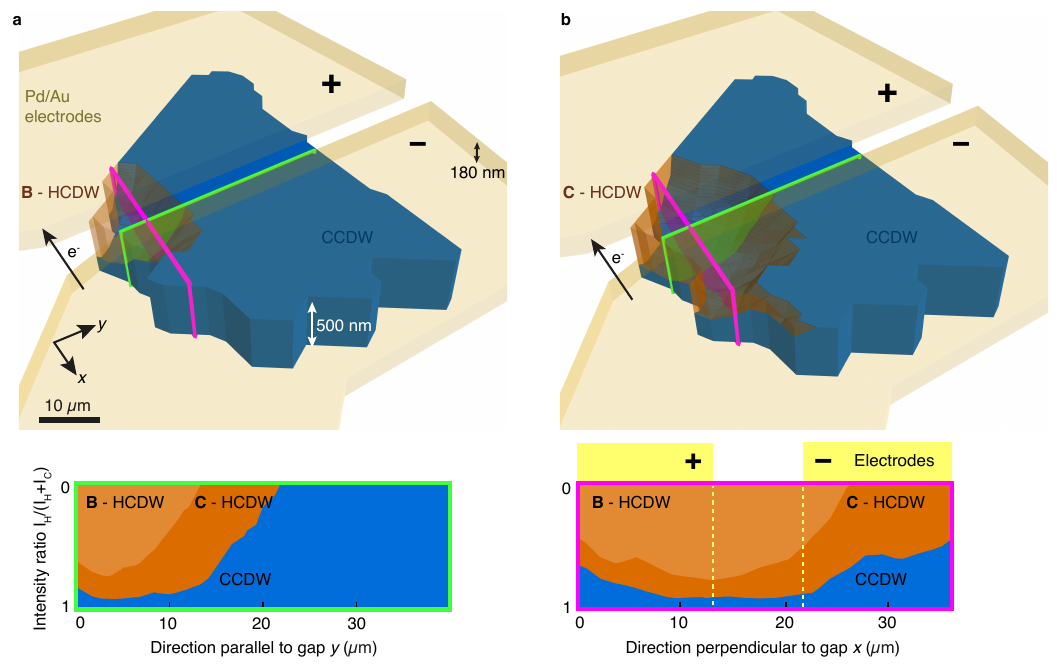}
     \caption{\textbf{Bulk electrical switching}.
        3D tomographic representation of the device in \textbf{a} the partially- \textbf{B} and \textbf{b} fully-switched state~\textbf{C}. The vertical dimension represents the normalized switching depth, defined by the ratio of the HCDW signal and the total~\mbox{(CCDW + HCDW)} intensity. 180-nm-thick Pd/Au electrodes (yellow) are on top of the 500-nm thick 1$T$-TaS$_2$ flake, where the CCDW region and the HCDW switching channel are depicted in blue and orange, respectively. Width and depth are not to scale. The arrow indicates the electron (e$^-$) flow upon application of the current pulse. Cuts through the flake parallel~(green) and perpendicular (pink) to the electrode gap are shown on the bottom left and right, respectively. Going from \textbf{B} to \textbf{C}, the in-gap HCDW order extends both laterally towards the $-$ electrode and in volume.}
    \label{fig_4}
\end{figure*} 

Next, we use the intensity shift from the CCDW to the HCDW peak to assess the switching depth. We take the ratio of the HCDW and the total \mbox{(HCDW $+$ CCDW)} signals as a proxy for the volume fraction of the switched layers. Since the Pd/Au electrodes are on top of the flake---thus, the highest current density is there---we also assume that the switching starts from the top. Figure~\ref{fig_4}~shows the respective 3D~tomographic representation of the device.
Clearly, the switching is not restricted to the surface but penetrates deep into the bulk of the 500-nm-thick flake. This observation is in agreement with previous reports on the importance of the out-of-plane layer reconfiguration in the HCDW state~\cite{Ma2016,Stahl2020,Hua2024}.
We take cuts parallel and perpendicular to the electrode gap to further characterize the bulk material switching: The HCDW region extends in volume going from the partially- \textbf{B} to the fully-switched state~\textbf{C}. Parallel to the gap the cut reveals that the switching starts at the edge of the flake that is the shortest path to ground. Moreover, from the cut perpendicular to the gap we see that the electrons flowing from the $-$ to the $+$ electrode do not symmetrically switch the intergap space. In addition, a fraction of $\approx10 \%$ remains unswitched between the electrodes. Though using XRD, we cannot extract information down to the single layer and consequently the remaining unswitched intergap layers are within the error of our technique. We also observe sizeable HCDW order induced under the electrodes that is not directly the shortest path to ground, hinting that not only the out-of-plane layer reordering~\cite{Ma2016,Stahl2020,Hua2024} but also strain and the out-of-plane conductivity~\cite{Svetin2017,Salgado2019} play an important role in the switching process.

\section*{Discussion}

We non-destructively acquire \textit{in operando} ``tomograms" of cryo-memory device switching, addressing a long-standing challenge for the engineering of phase-change memory devices~\cite{Yang2013,Sun2019}. Our approach allows us to identify the switched HCDW state along one edge of the flake and extending under the electrodes into the bulk of the material. This hints at charge injection as the initiator of the switching. The appearance of a current-induced conducting boundary channel at the flake edge / electrode interface aligns also with previous findings \cite{Devidas2024}. 

Finite element method simulations (Supplementary Information) indicate that the geometry of the device dictates the switching channel's location: The switching occurs where the electrode gap is narrowest, providing the shortest path to ground and resulting in the highest current density during pulse application. Once the channel has formed, the current primarily flows through it, and the switched volume only extends when the current is further increased, \textit{i.e.}, from state \mbox{\textbf{B} to \textbf{C}} (Fig.~\ref{fig_4}).
The Au~electrodes ensure high conductivity along the flake, rendering a conductivity gradient unlikely~\cite{Gilani2018}. The switching channel also does not seem to form along defects or impurities, as its size \mbox{($\approx20\times$ \SI{20}{\micro\meter\squared})} exceeds that of defects or domains~\cite{Stoneham1979,Gerasimenko2019}. Moreover, the lattice peak maps in Fig.~\ref{fig_2}a,d,g confirm that the flake remains intact during the switching process, ruling out structural fracture as the root cause. 

The switching region can also be located by mapping the vanishing dimer peak (Supplementary Information), associated with the breaking of the pairwise alignment of vdW layers \cite{Stahl2020}. The nature of the equilibrium CCDW state---whether it is a Mott or a band insulator---remains controversial and is found to be tied to the out-of-plane restacking \cite{Hua2024}. Here, we confirm control and local modification of the out-of-plane stacking via current pulses.

In- and out-of-plane strain between the flake and the substrate or electrodes may not only influence the stability of the HCDW state~\cite{Svetin2014, Vaskivskyi20152} but also the switching region's location. Studies on other phase-change materials show that local strain inhomogeneities can arise upon electrical triggering~\cite{Salev2024}. We also observe an out-of-plane lattice contraction localized exclusively in the intergap space of our \mbox{1$T$-TaS$_2$} device when a current pulse is applied (Supplementary Information). This creates local strain between the switched and unswitched areas. Thus, we identify strain propagation via the CDW domains as an essential factor contributing to the localized and bulk-nature of the switching, in close analogy to what occurs for ferroelectric devices~\cite{McWhan1985}. Since strain is  a long-range effect, it may also explain why the switching extends beneath the electrodes---an aspect not captured by the strain-free simulations.

Interestingly, the original CCDW intensity is higher in the region where the switching channel forms. Observed only for the CCDW but not the lattice peak (Fig.~\ref{fig_2}a,b), this indicates that structural differences, such as variations in flake thickness, are not responsible but rather that switching is potentially promoted by a particularly coherent (well-ordered) CDW state. Our results, therefore, underscore the importance of resolving electronic heterogeneities and looking into the role they play in the non-thermal switching process of 1$T$-TaS$_2$. 

Our 3D reconstruction of the phase change suggests optimized device designs using narrow flakes with electrodes positioned along the edges. Ultimately, we envision separate electrode pairs that allow for multiple switched regions and controlled cross talk between them, eventually enabling highly-efficient and fast logic operations.

Studying bulk switching using XRD with nano/micro-beams offers many opportunities. Direct imaging of filamentary paths arising from strain or different structural reconfiguration can also be applied to other memristive materials. So far these have mostly been investigated in thin films without contacts \cite{Singer2018}, in-plane with transmission electron microscopy, or indirectly and destructively using conductive atomic force microtomography~\cite{Bongjin2013,Celano2014,Toh2022,Wei2023}. Our non-destructive, 3D, \textit{in operando} approach represents a significant advance over these technologies, providing microscopic detail of the switching mechanism, thereby advancing the design of scalable next-generation memory technologies.

\section*{Methods}\label{sect-Methods} 
\textbf{Device preparation.}
1$T$-TaS$_2$ single-crystals were synthesized using chemical vapor transport with iodine as the transport agent leading to consistent crystal quality~\cite{Klanjsek2017}.
Flakes were mechanically exfoliated from bulk crystals using GelPak and excess flakes removed using Scotch tape such that a large, single flake could be deposited onto an oxidized Si wafer (oxide thickness:~\SI{280}{\nano \meter}) with predefined \mbox{Ti/Au}~alignment markers. The electrodes were written on the flake using electron beam lithography, followed by physical vapor deposition of \SI{20}{\nano\meter}~Pd and \SI{160}{\nano\meter}~Au. The investigated flake has lateral dimensions of about \mbox{50 $\times$ \SI{50}{\micro\meter\squared}} and a thickness of \SI{500}{\nano\meter}. The in-plane crystal orientation of the device was characterized at the Material Science beamline of Swiss Light Source synchrotron \cite{Willmott2013}, and it was then glued onto a Cu~sample holder using GE~varnish.\\

\textbf{Transport measurements and sample cooling.}
The device was mounted in a CryoVac $^4$He~cryostat with Kapton windows for the incident and scattered X-rays. The resistance was measured in a two-probe configuration on the flake, and converted to four-probes on the sample holder such that only the resistance of the Pd/Au electrodes on the flake is measured in series with the flake. The electrical setup consisted of a Keithley 6221 pulsed current source and a Keithley 2182 nano-voltmeter connected in delta mode. For electrical switching single, square-wave pulses were applied using the current source. The sample temperature was monitored using a Cernox thermometer mounted on the sample holder about \SI{2}{\centi\meter} from the device. A specific cooling procedure ensured that the first-order phase transition to the CCDW state was crossed in a controlled fashion. The following ramp rates were used: 300 to \SI{250}{\kelvin}:~\SI{5}{\kelvin}\slash \si{\minute}, 250 to \SI{200}{\kelvin}: \SI{2}{\kelvin}\slash \si{\minute}, 200~to~\SI{100}{\kelvin}: \mbox{\SI{1}{\kelvin}\slash \si{\minute}}, and 100 to \SI{6}{\kelvin}: \mbox{\SI{5}{\kelvin}\slash \si{\minute}}.\\

\textbf{X-ray diffraction and fluorescence. }
Spatially-resolved X-ray scattering was performed at the microXAS beamline of the Swiss Light Source~\cite{Grolimund2002}, allowing for simultaneous \textit{in situ} resistance measurements, as well as $\mu$XRD and $\mu$XRF imaging. The incidence angle of the X-rays with respect to the device was fixed to $\approx$\SI{25}{\degree}, resulting in a beam spot size \mbox{$\approx1.5\times$ \SI{2.5}{\micro\meter\squared} (v $\times$ h)} using  Kirkpatrick-Baez focusing mirrors. The nominal X-ray flux was adjusted to about \si{10^9}~photons/s. Considering that the X-rays were passing through about \SI{10}{\centi \meter} air before impinging on the sample, the effective incident flux on the sample was about \SI{15}{\percent} and \SI{10}{\percent} lower at 9.1 and \SI{12.0}{keV}, respectively. As the cryostat was mounted on translation stages without rotational degree of freedom, to take $\mu$XRD spatial maps we scanned through reciprocal space by changing the incoming \mbox{X-rays} energy between \SI{9.1} and \SI{12.0}{keV} using the undulator and monochromator. The diffraction and fluorescence signals were recorded by Eiger X 4M and VIAMP-KC detectors, respectively.
To locate the device the $\mu$XRF signal at the Au (\SI{11.9}{\kilo\electronvolt}) and Ta (\SI{9.88}{\kilo\electronvolt}) $L$-edges was used (Supplementary Information).

\section*{Data availability}
Data supporting the figures and other findings of this study are available from the corresponding authors upon request.

\begin{acknowledgments}

The authors acknowledge the Paul Scherrer Institute, Villigen, Switzerland for provision of synchrotron radiation at the microXAS and Material Science beamlines of the Swiss Light Source. We thank P.~Sutar for synthesizing the \mbox{1$T$-TaS$_2$} crystals and acknowledge technical support from the PSI~PICO operations team. Furthermore, we are grateful to B.~Meyer, M.~Birri and S.~Stutz for technical support before and during the beamtimes, as well as fruitful discussions with N.~Taufertshöfer. This research was funded by the Swiss National Science Foundation~(SNSF) and the Slovenian Research And Innovation Agency~(ARIS) as a part of the WEAVE framework Grant Number~213148  (ARIS project N1-0290). R.V., A.M. and D.M. thank ARIS for funding the research program P1-0400, and D.M. thanks ARIS for funding the research program J7-3146. B.L. and M.T. thank ARIS for funding the research program P2-0415. C.B., W.H., A.G.M. and G.A. acknowledge funding from the European Research Council under the European Union’s Horizon 2020 research and innovation program, within Grant Agreement 810451 (HERO). N.H. received funding from the European Union’s Horizon 2020 research and innovation program under the Marie Sklodowska-Curie Grant Agreement 884104 \mbox{(PSI-FELLOW-III-3i)}.

\end{acknowledgments}

\section*{Author contributions}
C.B., D.M. and S.G. conceived the project with input from G.A. and Y.E. C.B. and D.K. fabricated the device with input from D.S. C.B., A.M., N.H., S.-W.H., F.D. and S.G. characterized the device. C.B., N.H., D.F.S., R.V., F.D., D.G. and S.G. prepared the experiment. C.B., N.H., D.F.S., W.H., H.G.B., R.V., F.D., S.-W.H., A.G.M., and S.G. carried out the experiment with input from D.G. 
C.B., W.H., H.G.B., and N.H. analyzed the data.
B.L. conducted the simulations with input from M.T., C.B., A.M.,  D.M. and S.G.
C.B., N.H., D.M. and S.G. wrote the manuscript with input from all co-authors.

\section*{Competing interests}
The authors declare no competing interest.


\clearpage
\begin{table*}[ht]
\centering
\begin{tabular}{|c||c|c|c|}
\hline \textbf{Peak} & \textbf{Literature}  & \textbf{Experiment} & \textbf{Exp. uncertainty}\\
\hline\hline Lattice & $(0,1,3)$ & $(0.00,1.00,3.00)$ &  $(3.7e^{-4},3.7e^{-5},3.6e^{-3})$\\
\hline CCDW & $(0.08,1.23,3.80)$ & $(0.08,1.23,3.83)$  & $(1.0e^{-3}, 7.7e^{-4},3.4e^{-2})$\\
\hline HCDW & $(0.07,1.24,3.67)$ & $(0.07,1.24,3.68)$  &  $( 3.5e^{-4},5.8e^{-4},4.2e^{-2})$\\
\hline
\end{tabular}
\caption{\textbf{Peak positions from electrically- and optically-switched experiments}. Reciprocal space coordinates \mbox{(\textit{h, k, l})} in r.l.u. of the measured structural (013), CCDW and e-HCDW peaks, as well as the respective literature values for the CCDW~\cite{Lauhle2015} and the o-HCDW~\cite{Stahl2020} signals. The experimental position and uncertainty are determined from fitting Lorentzian peak shapes after the reciprocal space reconstruction.}
\label{table:Table1}
\end{table*}

\vspace{\baselineskip}

\bibliography{Main/Manuscript}
\clearpage


\title{Supplementary information \\
\textit{Imaging of electrically controlled van der Waals layer stacking in 1$T$-TaS$_2$}}

\author{Corinna Burri}
\affiliation{PSI Center for Photon Science, Paul Scherrer Institute, 5232 Villigen PSI, Switzerland}
\affiliation{Laboratory for Solid State Physics and Quantum Center, ETH Zurich, 8093 Zurich, Switzerland}

\author{Nelson Hua}
\author{Dario Ferreira Sanchez} 
\affiliation{PSI Center for Photon Science, Paul Scherrer Institute, 5232 Villigen PSI, Switzerland}

\author{Wenxiang Hu}
\author{Henry~G.~Bell}
\affiliation{PSI Center for Photon Science, Paul Scherrer Institute, 5232 Villigen PSI, Switzerland}
\affiliation{Laboratory for Solid State Physics and Quantum Center, ETH Zurich, 8093 Zurich, Switzerland}

\author{Rok~Venturini}
\affiliation{PSI Center for Photon Science, Paul Scherrer Institute, 5232 Villigen PSI, Switzerland}
\affiliation{Department of Complex Matter, Jozef Stefan Institute, 1000 Ljubljana, Slovenia}

\author{Shih-Wen~Huang} 
\affiliation{PSI Center for Photon Science, Paul Scherrer Institute, 5232 Villigen PSI, Switzerland}

\author{Aidan G. McConnell}
\author{Faris~Dizdarević}
\affiliation{PSI Center for Photon Science, Paul Scherrer Institute, 5232 Villigen PSI, Switzerland}
\affiliation{Laboratory for Solid State Physics and Quantum Center, ETH Zurich, 8093 Zurich, Switzerland}

\author{Anže~Mraz}
\affiliation{Department of Complex Matter, Jozef Stefan Institute, 1000 Ljubljana, Slovenia}
\affiliation{CENN Nanocenter, 1000 Ljubljana, Slovenia}
\author{Damjan~Svetin}
\affiliation{Department of Complex Matter, Jozef Stefan Institute, 1000 Ljubljana, Slovenia}

\author{Benjamin Lipovšek}
\author{Marko Topič}
\affiliation{Faculty for Electrical Engineering, University of Ljubljana, 1000 Ljubljana, Slovenia}

\author{Dimitrios~Kazazis} 
\affiliation{PSI Center for Photon Science, Paul Scherrer Institute, 5232 Villigen PSI, Switzerland}

\author{Gabriel~Aeppli}
\affiliation{PSI Center for Photon Science, Paul Scherrer Institute, 5232 Villigen PSI, Switzerland}
\affiliation{Laboratory for Solid State Physics and Quantum Center, ETH Zurich, 8093 Zurich, Switzerland}
\affiliation{Institute of Physics, EPF Lausanne, 1015 Lausanne, Switzerland}  

\author{Daniel Grolimund}
\affiliation{PSI Center for Photon Science, Paul Scherrer Institute, 5232 Villigen PSI, Switzerland}

\author{Yasin Ekinci} 
\affiliation{PSI Center for Photon Science, Paul Scherrer Institute, 5232 Villigen PSI, Switzerland}

\author{Dragan Mihailovic}
\email{dragan.mihailovic@ijs.si}
\affiliation{Department of Complex Matter, Jozef Stefan Institute, 1000 Ljubljana, Slovenia}
\affiliation{CENN Nanocenter, 1000 Ljubljana, Slovenia}
\affiliation{Faculty of Mathematics and Physics, University of Ljubljana, 1000 Ljubljana, Slovenia}

\author{Simon Gerber}
\email{simon.gerber@psi.ch}
\affiliation{PSI Center for Photon Science, Paul Scherrer Institute, 5232 Villigen PSI, Switzerland}

\date{\today}

\maketitle



\section{Fluorescence}
\begin{figure*}[htpb]
  \centering
    \includegraphics[width=0.75\linewidth, trim=0cm 0cm 0cm 0cm, clip=true]{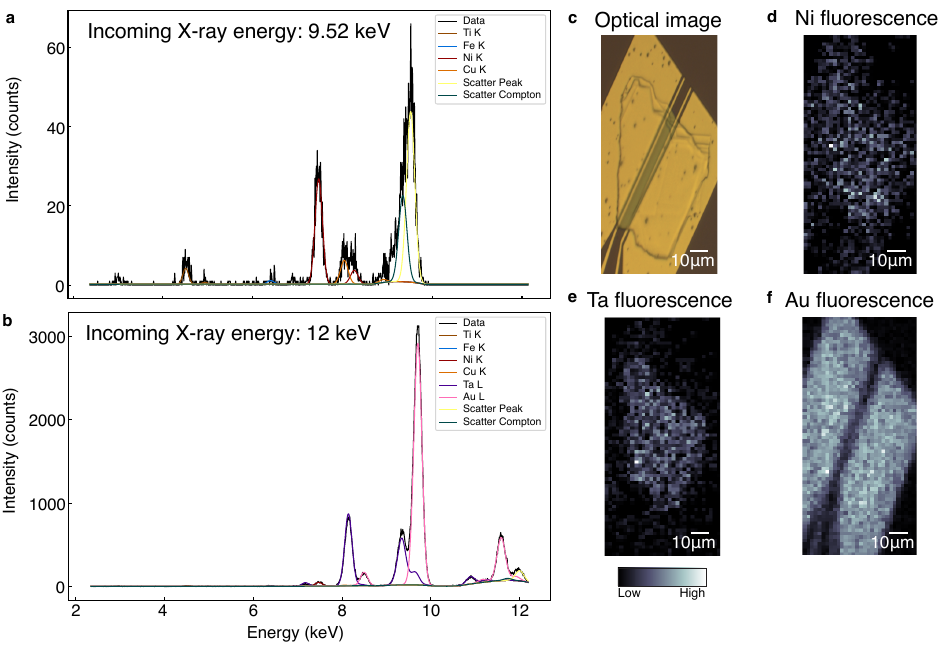}
    \caption{\textbf{Elements in the 1\textit{T}-TaS$_2$ device detected by X-ray fluorescence}. 
     \textbf{a} Total fluorescence intensity measured using an incoming X-ray energy of \SI{9.52}{\kilo\electronvolt}, below the Ta \textit{L}-edge \cite{Thompson2001}. Peak fitting traces elements such as Ti, Fe, Ni and Cu on the device and sample holder (via the respective \textit{K}-edges), as well as elastic and Compton scattering.
     \textbf{b} Respective fluorescence signal measured at \SI{12}{\kilo\electronvolt}, above the Ta and Au \textit{L}-edge \cite{Thompson2001}, where in addition to the elements observed in \textbf{a}, also Ta and Au are detected.
     \textbf{c} Optical image of the device. \textbf{d} Ni signal mapped spatially across the device showing traces along the electrodes and the flake. \textbf{e,f} The Ta and Au signals outline the flake and electrodes, respectively. The minimal intensity in \textbf{d}-\textbf{f} is 0~counts, and the maximal intensity is $6.7$ counts (Ni), $9.6$ counts (Ta), and $67$ counts (Au), measured using an integration time of \SI{100}{\milli\second}.
     }
    \label{fig: S10}
\end{figure*}

X-ray fluorescence of Ta and Au was used to locate the device and set the range of the spatially-resolved scans. The signal was captured by VIAMP-KC detectors. Even small traces of elements such as Cu from the sample holder, Ti from the binary markers on the substrate, and Ni traces under the electrodes were detected (Fig.~\ref{fig: S10}). Fluorescence data were fitted using the PyMCA libraries~\cite{Sole2007}.

\section{X-ray diffraction data processing}

\subsection{Experimental geometry constraint and reciprocal space reconstruction}
In the following, we describe how we assigned pixels on the detector at each X-ray energy (\textit{$x_d$, $y_d$, E}) to reciprocal space coordinates (\textit{h, k, l}).

The known parameters in this experiment are the energy of the X-rays, the orientation of the detector which was perpendicular to the X-ray beam and further detector specifications of the Eiger X 4M detector. In addition, we know the optimal pixel and energy of scattering reflections of three accessible lattice peaks (004), (013), (014) and four commensurate CCDW peaks on the detector and also their theoretical (\textit{h, k, l}) parameters \cite{Lauhle2015}, leading to seven known quantities. The measured peak positions were determined by fitting a Lorentzian to the projected intensities. The CCDW peaks are elongated along the out-of-plane direction, therefore only the in-plane (\textit{h, k}) dimensions were fitted.  
The unknown parameters in this setup were the relative position of the detector with respect to the sample (\textit{x,~y,~z}) which was roughly measured and used as an initial guess, the sample orientation parameterized with angles ($\alpha, \phi, \chi$) and the lattice constants of the flake (\textit{a, b, c}). If we parametrize the Eiger X 4M orientation with Cartesian coordinates (\textit{x,~y,~z}) where (\textit{x, y}) point along the \textit{x, y} pixel directions and \textit{z} perpendicular to the detector surface, the flake orientation can be expressed in this coordinate system using angles ($\alpha, \phi, \chi$). $\alpha$ denotes the angle between the surface normal of the detector and the surface normal of the flake, $\phi$ is the azimuthal angle and $\chi$ the tilt angle. Since there were only seven measured parameters (three lattice and four CCDW reflections), we took the in-plane lattice parameters $a=b=\SI{3.32}{\angstrom}$ as known \cite{JELLINEK19629} and only the out-of-plane lattice parameter \textit{c} as unknown.

As a next step, a cost function was defined that includes all known and unknown parameters, and with which the full experimental geometry was described as

\begin{equation}
F(\Vec{x},\Vec{y})=\sum_{i=1}^{7} \left|\Vec{Q_i}(\Vec{x},\Vec{y})-\Vec{Q_i}^{\mathrm{theory}}\right|^2,
\end{equation}
where $\vec{x}$ are the unknown parameters, $\vec{y}$ are the known parameters and the two-norm is used.

With this function, the reciprocal space coordinates of our measured peaks was compared to the theoretically known position \cite{Lauhle2015, Stahl2020}. The cost function was minimized with respect to $\Vec{x}$ using a standard minimization algorithm from the SciPy optimization python library \cite{Virtanen2020} which finds the experimental conditions $(\alpha, \phi, \chi, x, y, z, c)=(\SI{25.61}{\degree}, \SI{-86.42}{\degree}, \SI{0.57}{\degree}, \SI{-0.14}{\meter}, \SI{-0.06}{\meter}, \SI{0.0}{\meter}, \SI{0.6}{\nano\meter})$ of the unknown parameters that matches the experimental results. The resulting error table is shown in Tab.~\ref{table:Table2} with small errors, meaning that there is good agreement between the measured peaks and the known structure of the material.

\begin{table*}[htpb]
\begin{center}
\begin{tabular}{|c||c|c|c||}
\hline Peak & Literature (r.l.u) & Experiment (r.l.u) & Two-norm error (pixels) \\
\hline \hline (013) CCDW1 & $(0.08, 1.23, 2.80)$ & $(0.08, 1.23 ,2.86)$ & 1.5 \\
\hline (013) CCDW2 & $(-0.08, 0.77, 3.20)$ & $(-0.08, 0.77, 3.14)$ & 0.7 \\
\hline (014) CCDW1 & $(0.08,1.23,3.80)$ & $(0.07,1.24,3.68)$ & 1.2 \\
\hline (014) CCDW2 & $(-0.08,0.77,4.20)$ & $(-0.08,0.77,4.16)$ & 0.7 \\
\hline (013) lattice & $(0, 1, 3)$ & $(0.0, 1.0, 3.0)$ & 3.5 \\
\hline (004) lattice & $(0,0,4)$ & $(0.0, 0.0, 4.0)$ & 5.0 \\
\hline (014) lattice & $(0,1,4)$ & $(0.0, 1.0, 4.0)$ & 2.3 \\
\hline
\end{tabular}
\caption{\textbf{Error table of the measured peaks}. Lattice and nearby CCDW peaks detected in the experiment are listed with their index from the literature \cite{Lauhle2015,Stahl2020}, the calculated (\textit{h, k, l}) positions determined from the experimental geometry and the error from the peak fits. The latter is given in pixel and calculated by taking the norm of the absolute error divided by the reciprocal space pixel resolution of $1.03 \times 10^{-3}$ r.l.u.}
\label{table:Table2}
\end{center}
\end{table*}

Once this experimental geometry was calculated, the real space position with respect to the sample of all pixels on the detector were defined. The reciprocal space coordinates (\textit{h, k, l}) were assigned to the pixels on the detector at every X-ray energy using the Laue equation. We note that the pixel size of the Eiger X 4M detector is \si{75} $\times$ \SI{75}{\micro\meter\squared} whereas the lateral size of the flake was about \si{50} $\times$ \SI{50}{\micro\meter\squared}. Hence, any movement of the diffraction pattern due to rastering over the sample was still within 1 pixel of the detector and can be neglected.

\begin{figure*}[htpb]
  \centering
    \includegraphics[width=0.75\linewidth, trim=0cm 0cm 0cm 0cm, clip=true]{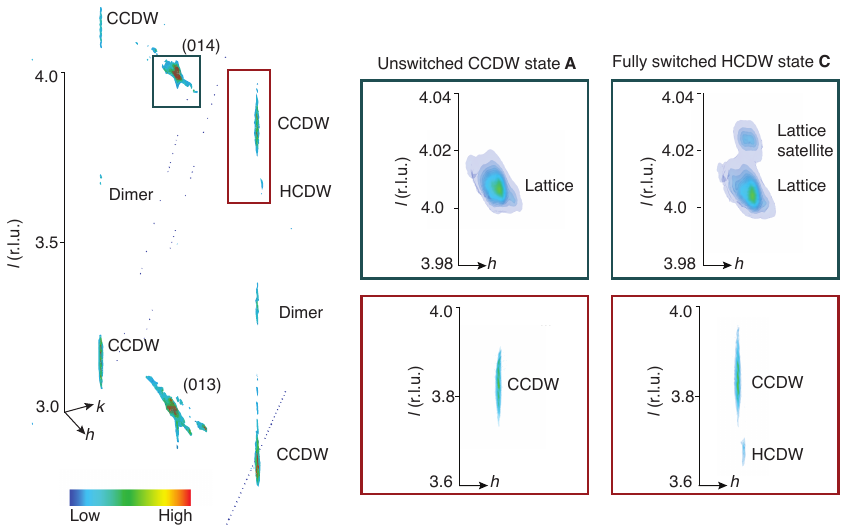}
    \caption{\textbf{3D reciprocal space reconstruction of the fully-switched state}. Shown on the left is the reconstruction of the fully-switched state \textbf{C} summed from all pixels on the flake (no spatial resolution), including the (013) and (014) lattice peaks, two CCDW peaks near each of them, a HCDW peak near the (014) peak and dimer peaks. This is a volumetric plot from the intensities of the peaks showing the general shape of the different peaks. The reconstruction is done for the measured energies between 9.2 and $\SI{12}{\kilo\electronvolt}$, which is why only certain peaks are observed. The blue pixel lines are due to dead pixels of the detector. On the right are extracts around the (014) lattice peak and a corresponding CCDW peak, showing the appearance of the HCDW peak and also a new lattice satellite when comparing the unswitched \textbf{A} and the fully-switched state \textbf{C}.
 }
    \label{fig: S1}
\end{figure*} 

The conversion from (\textit{$x_d$, $y_d$, E}) to (\textit{h, k, l}) leads to a curved surface in reciprocal space which was probed by the flat 2D detector in real space. Putting all the curved surfaces of the measured energies together, allowed us to reconstruct the 3D shape of the lattice, CCDW, HCDW and dimer peaks. Figure~\ref{fig: S1} shows this 3D reciprocal space representation measured in the fully-switched state~\textbf{C}. This is a volumetric plot using the intensities of the measured peaks, giving an overview over the peak shapes and position. Importantly, this is not a spatially-resolved 3D~reciprocal image, it is the reconstruction coming from the sum of all pixels measured at the different energies, therefore we observe all peaks in the same plot. Next to the overview plot are extracts of reciprocal space around the lattice and CCDW peak in the unswitched~\textbf{A} and in the fully-switched state~\textbf{C}. We see that in the fully-switched state a HCDW peak appeared below the CCDW peak, as well as an additional lattice satellite (Sect.~\ref{Outofplane}). For the spatially-resolved intensity maps of the device, we used such 3D reconstruction of every pixel. The peak positions were determined from the 3D~reconstruction by summing along two reciprocal space directions to obtain a 1D intensity plot along the remaining reciprocal space direction and fitting a Lorentzian function to the peak.

\subsection{Gold ring correction}
\begin{figure*}[htpb]
  \centering
    \includegraphics[width=\linewidth, trim=0cm 0cm 0cm 0cm, clip=true]{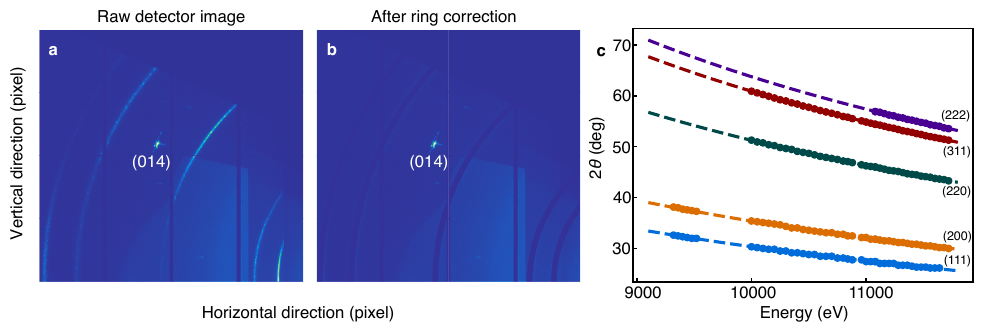}
    \caption{\textbf{Gold ring correction}. 
    \textbf{a} Exemplary raw detector image taken at \SI{11.12}{\kilo\electronvolt} and showing five rings coming from the Au electrode material as well as the (014) lattice peak of the 1$T$-TaS$_2$ flake. \textbf{b} The same detector image after fitting the gold rings and replacing them with an average background. \textbf{c} Fitting of the five gold rings observed in the experiment (data points), corresponding to the (111), (200), (220), (311) and (222) Au reflections from the literature \cite{Graf2017} (dashed lines).
    }
   \label{fig: S2} 
\end{figure*}

We observed five ring-like features in the raw diffraction patterns (Fig.~\ref{fig: S2}a). After the detector position with respect to the sample was determined, the \si{2\theta} angle of these rings, \textit{i.e.} the radius $r$, at a certain energy of the incoming X-rays was determined using Bragg's law
\begin{equation}
\sin(2\theta)=\frac{n\lambda}{2d}=\frac{\lambda}{2k},\,\tan(\theta)=\frac{r}{d},
\end{equation}

where $\lambda$ is the  wavelength of the incoming X-rays, $d$~the spacing between the crystal planes, $n$ an integer for the diffraction order and $k$ the wavenumber.
Comparison with literature showed that the radii of the rings matches Au with Miller indices (111), (200), (220), (311) and (222) \cite{Graf2017} (Fig.~\ref{fig: S2}c). The electrodes are made from amorphous gold, therefore we observe these rings when spatially scanning over the device. Taking a region of interest (ROI) with radius $r \in [r_{\mathrm{ring}}-0.0018, r_{\mathrm{ring}}+0.0018$] (in m) corresponding to the thickness of the gold rings, allowed us to collect and sum up the intensity of all detector pixels within this ROI. We did this procedure for every pixel ($x_d$ and $y_d$) on the sample which gave us spatially-resolved intensity maps outlining the electrodes of the device. We used these electrode images for aligning the spatial scans at the different X-ray energies. 

After confirming that no peak of interest cut through the gold rings, we replaced them for the 3D reciprocal space reconstruction by an average background (Fig.~\ref{fig: S2}b). This was simply done for better visibility of the peak shapes in the reconstruction, but did not affect the reconstruction itself.

\subsection{Spatially-resolved maps}
In the following, we describe how we obtain the spatially-resolved intensity maps, as well as the spatially-resolved projections in reciprocal space.
For each state of the device, \textit{i.e.} the room-temperature, as well as the unswitched \textbf{A}, partially- \textbf{B} and fully-switched states~\textbf{C} at the base temperature of 6~K, we measured spatial maps at different energies. As a next step, we performed a spatial jitter correction. That is, we spatially aligned the scans at different energies and for the different states. We did this by choosing a gold ring which shows clear outlines of the electrodes in the spatial maps.
For each state, we first chose one reference spatial map~$I_{ref}$ measured at a certain X-ray energy. We then aligned all other spatial maps measured at different energies to this reference map. Due to the finite scan range, we confined a possible shift in the \textit{x} and \textit{y} spatial directions to $\delta x \in[-10,10]$ and $\delta y \in[-20,20]$ (in pixel) and found $(\delta x, \delta y)$ such that

\begin{equation}
 F(\delta x, \delta y)=\sum_{i,j}\left[I_{r e f}\left(x_i, y_j\right)-I\left(x_i+\delta x, y_j+\delta y\right)\right]^2,   
\end{equation}
summing over the pixel indices $i, j$, was minimized. The result shows in general good alignment within ±1 pixel. But there were a few scans at certain energies which still had to be manually tweaked to improve the alignment. 

After having the scans at different energies for every state spatially aligned and a 3D reciprocal space reconstruction for each pixel performed, we defined 3D ROIs in reciprocal space around the peaks of interest. Since the CCDW and dimer peaks are elongated along \textit{l}, as has been reported before \cite{Stahl2020}, we took cylindrical ROIs around these peaks. For the lattice and HCDW peaks which are better defined along the out-of-plane direction, we took spherical ROIs. We computed the signal within these ROIs by summing up the intensities.
For every spatial map at every energy, we set a spatial 1.5 $\times$ \SI{1.5}{\micro\meter\squared} (room temperature) and \si{1} $\times$ \SI{1}{\micro\meter\squared} (\SI{6}{\kelvin}) grid according to the minimal step size of the scan. Then, we assigned intensity values to the grid by considering that if the closest scan to a respective grid pixel is $>$ 2~pixel, we set it to zero and otherwise we consider the data within the predefined ROI in reciprocal space. For the spatially-resolved intensity maps, such as in Fig.~2 of the main text, the intensity in the spatial pixels was therefore the sum of the intensities within the ROI. We note that with a minimal step size in the $x$ and $y$ direction of \SI{1.5}{\micro\meter} at room temperature and \SI{1}{\micro\meter} at the base temperature, we were in fact oversampling given the X-ray spot size of \mbox{1.5 $\times$ \SI{2.5}{\micro\meter\squared}} (v $\times$ h).

After identifying the region which switched to the HCDW state from the spatial maps, we could separate the device into different spatial regions to extract the spatially-resolved 2D reciprocal space projections (Fig.~3 of the main text) and calculate the ratio of the HCDW and the CCDW peak intensity. The spatial region near the intergap space (light purple) corresponds to a intensity ratio $0.5$ and that region in the switching channel (dark purple) corresponds to a ratio of $5$. 
Then, we summed up the 3D reciprocal space reconstructions of the pixels within these regions. For the projected maps, such as Fig.~3 of the main text, we set a grid in reciprocal space and interpolate. From this, we obtain the 2D projections by summing along one reciprocal space axis and dividing by the number of pixels in the spatial regions.

\section{Room-temperature spatial maps}
\begin{figure}[tpb]
  \centering
    \includegraphics[width=0.57\linewidth, trim=0cm 0cm 0cm 0cm, clip=true]{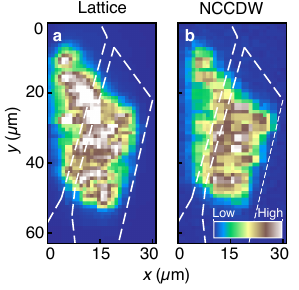}
    \caption{\textbf{Room-temperature spatially-resolved maps}. 
    \textbf{a}, \textbf{b} show the (013) lattice and NCCDW peak intensities, respectively. The minimal intensity is 0 counts and the maxima are $6.0 \cdot 10^4$ counts (013) and $1.0 \cdot 10^3$ counts (NCCDW) for an integration time of~\SI{100}{\milli\second}. Dashed lines outline the electrodes. The step size was \SI{1.5}{\micro\meter}.
}
    \label{fig: S3}
\end{figure}

At the beginning of the experiments, the device was mounted at room temperature and spatial scans at different X-ray energies were measured to check the alignment, as well as to record the lattice and NCCDW signals. We took spatially-resolved scans at a few energies, corresponding to only selected cuts through reciprocal space. Hence, we only observed a few peaks with reduced statistics, compared to the low-temperature measurements. This resulted in a less accurate 3D reconstruction. Nevertheless, we took 3D ROIs around the (013) lattice peak and a respective NCCDW peak and plotted the corresponding spatial intensity map which both reproduce the general shape of the flake (Fig.~\ref{fig: S3}).

\section{Dimers and phase switching}
\begin{figure}[tpb]
  \centering
    \includegraphics[width=0.75\linewidth, trim=0cm 0cm 0cm 0cm, clip=true]{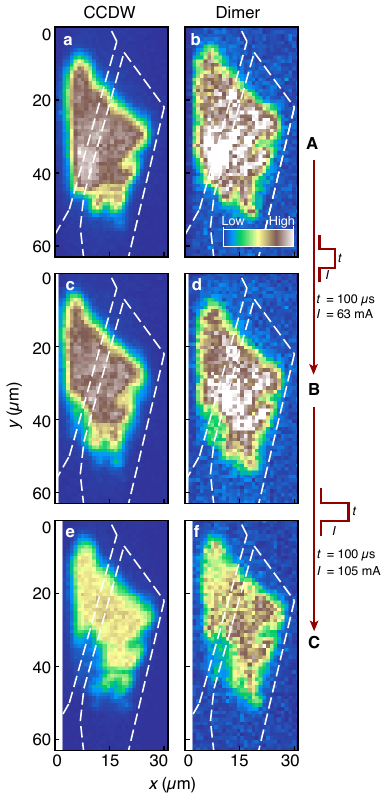}
    \caption{\textbf{Vanishing dimers in the switching channel}. 
    \mbox{\textbf{a},\textbf{b}} Spatially-resolved intensity of the CCDW and dimer peaks in the unswitched state \textbf{A}, respectively. \textbf{c},\textbf{d} and \mbox{\textbf{e},\textbf{f}} show the maps in the partially- \textbf{B} and fully-switched state~\textbf{C}, respectively. The dimer peak vanishes in the latter, \textit{i.e.} the switching channel can also be seen from the dimer signal. The minimal intensity is 0 counts and the maxima are $1.5 \cdot 10^3$ counts (CCDW) and $1.0 \cdot 10^2$ counts (dimer) for an integration time of \SI{100}{\milli\second}. Dashed lines outline the electrodes. The step size was \SI{1}{\micro\meter}.}
    \label{fig: S4}
\end{figure}

As discussed in \cite{Stahl2020, Wang2020}, dimer peaks associated with a doubling of the real-space unit cell in the out-of-plane direction were observed in the CCDW state and vanish upon switching to the o-HCDW state \cite{Stahl2020}. We also observed dimer peaks in our measurements (Fig.~\ref{fig: S1}) and took 3D ROIs around those in reciprocal space. We recorded only a few energy scans cutting through the dimer peaks, meaning that the statistics was reduced compared to other peaks. Figure~\ref{fig: S4} shows the spatially-resolved intensity maps of the CCDW and dimer peak in the unswitched \textbf{A}, the partially- \textbf{B} and the fully-switched state \textbf{C}. We see that both peak intensities vanish in the bottom left part of the flake in the fully- switched state. Hence, we observe the same vanishing dimer peaks in the e-HCDW state in the intergap space as reported for the o-HCDW state, meaning for both there is also collapse of dimerization. Indirectly, observed as a local doubling of the out-of-plane dispersion, this has also been reported in a recent $\mu$ARPES study \cite{Nitzav2024}. As addressed in the main text, we observed that the intensity of the CCDW and dimer peaks in the unswitched state was the highest in the bottom left part of the flake that switches to the HCDW state. This supports the notion that the location of the switching channel may not only be dictated by the geometry of the device, but also the most homogeneous (dimerized) CDW order.

\section{Switching-induced out-of-plane lattice contraction}\label{Outofplane}
\begin{figure}[tpb]
  \centering
    \includegraphics[width=0.7\linewidth, trim=0cm 0cm 0cm 0cm, clip=true]{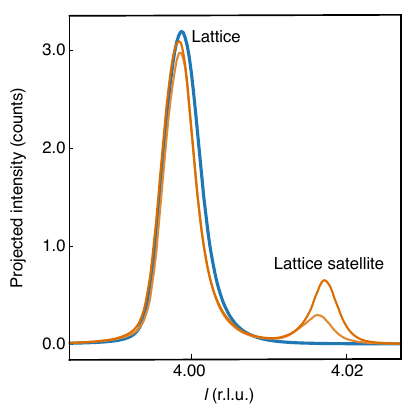}
    \caption{\textbf{Out-of-plane lattice contraction}. 
    Out-of-plane projection of the (004) lattice peak in the unswitched (\textbf{A},~blue), the partially- (\textbf{B}, light orange) and  the fully-switched state (\textbf{C}, dark orange). This measurement integrates over the entire flake. Upon switching intensity shifts from the (004) peak to the satellite at higher \textit{l} values, corresponding to an out-of-plane lattice contraction of \SI{0.5}{\percent}.
 }
    \label{fig: S5}
\end{figure}

As reported before for the o-HCDW state \cite{Vaskivskyi2024}, we also observed a change in the out-of-plane lattice constant in the e-HCDW state. Figure~\ref{fig: S5} shows the 1D projection onto the out-of-plane reciprocal space coordinate \textit{l} of the (004) lattice peak. The signal was measured on the entire flake (no spatial resolution) in the unswitched~\textbf{A}, the partially- \textbf{B} and the fully-switched state \textbf{C}. The high intensity peak is the (004) lattice peak reflection. We note that because this measurement is not spatially-resolved, intensity (from the unswitched portions of the flake) still shows up there even in the \textbf{C} state. Clearly, in the \textbf{B} and \textbf{C} states a satellite lattice peak emerges from the HCDW state in the switching channel. Thus, we observed that for the e-HCDW state the out-of-plane lattice peak position shifts to a higher reciprocal space value, meaning that the lattice contracts by \SI{0.5}{\percent}. The out-of-plane lattice constant difference between the switched and unswitched region leads to strain along the border which we speculate facilitates the switching in-plane but also in volume. Similarly, the \textit{c} lattice constant contracts between the NCCDW and CCDW states \cite{Givens1977,Sezerman1980,Guy1985,Wang2020}. 

\begin{figure}[tpb]
  \centering
    \includegraphics[width=0.75\linewidth, trim=0cm 0cm 0cm 0cm, clip=true]{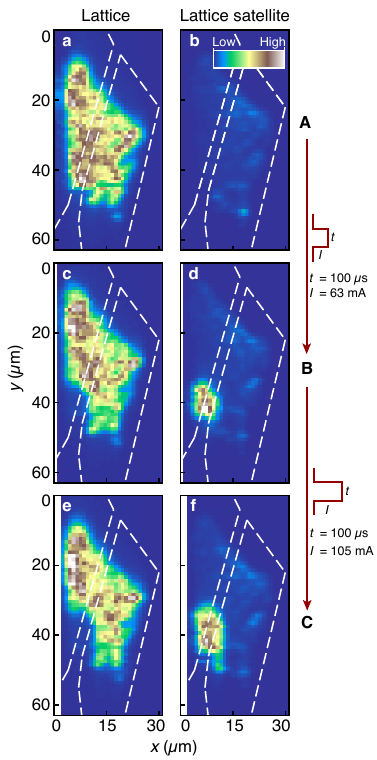}
    \caption{\textbf{Spatially-resolved out-of-plane lattice contraction}. 
    \textbf{a},\textbf{b} Intensity map at the (004) lattice peak and satellite position in the unswitched state \textbf{A}, respectively. \mbox{\textbf{c},\textbf{d}} and \textbf{e},\textbf{f} show the respective maps in the \mbox{partially- \textbf{B}} and fully-switched state \textbf{C}, respectively. As for the CCDW and HCDW signal, the main lattice peak vanishes in the bottom left part of the flake and the satellite lattice peak is observed instead, showing that the switching channel can also be observed via the out-of-plane lattice contraction. Color scales show a minimal intensity of 0~counts and maxima of $1.0 \cdot 10^3$ counts for an integration time of~\SI{100}{\milli\second} for both columns. Dashed lines outline the electrodes. The step size was \SI{1}{\micro\meter}.}
    \label{fig: S6}
\end{figure}

We set two ROIs around the lattice signals in reciprocal space: $l \in(3.99, 4.01)$ and $(4.015, 4.035)$ r.l.u. for the main and satellite peak, respectively, to obtain spatial maps by summing up the intensity within these ROIs (Fig.~\ref{fig: S6}). Also here we observe that the HCDW state switching channel appeared at the bottom left corner of the flake. Thus, not only the electronic CDW structure is modified in the e-HCDW state but also the out-of-plane lattice contracts in the non-volatile switching channel of the device. In the following, we show the agreement of the HCDW state switching channel observed from the out-of-plane lattice contraction and the HCDW peak. 

\begin{figure*}[htpb]
  \centering
    \includegraphics[width=0.75\linewidth, trim=0cm 0cm 0cm 0cm, clip=true]{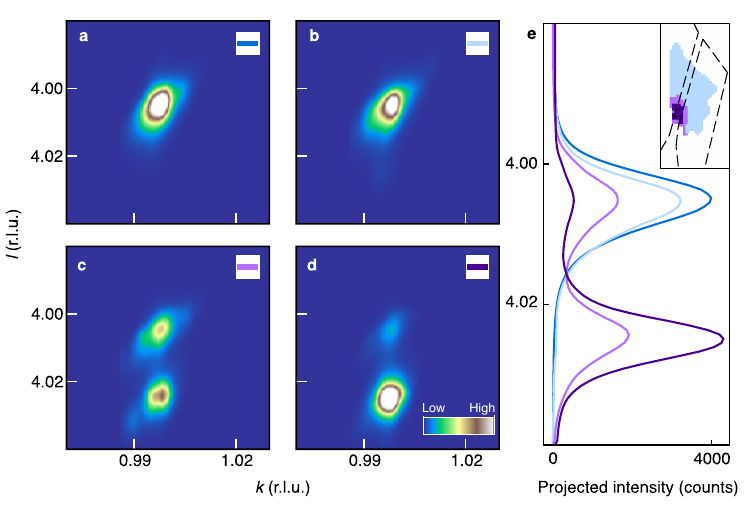}
    \caption{\textbf{Momentum- and real-space structure of the lattice contraction}. 
     \textbf{a} 2D (\textit{k, l}) reciprocal space projection of the lattice peak in the unswitched state (\textbf{A}, dark blue). \textbf{b-d} Respective measurement in the fully-switched state \textbf{C} in the light blue, as well as light and dark purple region (inset of e), respectively. The 2D projected intensities have been normalized by the size of the respective spatial region. \textbf{e} Out-of-plane projection of the data shown in a-d. The Inset shows the color-coded regions on the flake. Dashed lines in the inset outline the electrodes.}
    \label{fig: S7}
\end{figure*}

\begin{figure}[htpb]
  \centering
    \includegraphics[width=\linewidth, trim=0cm 0cm 0cm 0cm, clip=true]{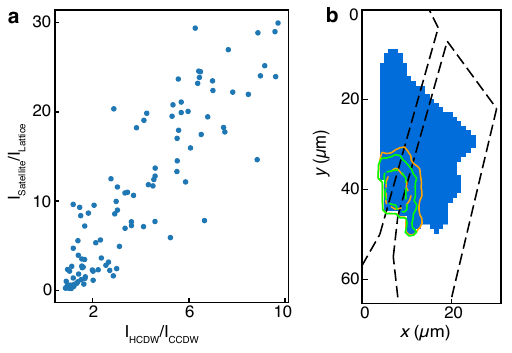}
    \caption{\textbf{Correlation between lattice contraction and HCDW peak appearance}. 
     \textbf{a} Linear correlation among the ratio of the out-of-plane lattice signals and the HCDW/CCDW peak intensity. \textbf{b} The blue region indicates the shape of the flake. The spatial switching region as identified by the lattice contraction and HCDW appearance is indicated in orange and green, respectively. Dashed lines outline the electrodes.}
    \label{fig: S8}
\end{figure}

As done for Fig.~3 of the main text, we define different spatial regions of the device: outside of the switching channel (light blue), near the switching channel (light purple) and in-between the electrodes (dark purple). We then map out the 2D reciprocal space projections of the (004) lattice peak (Fig.~\ref{fig: S7}). Unlike the CCDW peak which shows a characteristic elongation, the lattice peak is sharp also in the out-of-plane direction. The correlation of the ratio of the projected out-of-plane lattice intensities of the main and satellite signal, $I_{\rm satellite}/I_{\rm main}$, and that of the HCDW and CCDW peak intensity, $I_{\rm HCDW}/I_{\rm CCDW}$, is shown in Fig.~\ref{fig: S8}a.
Figure~\ref{fig: S8}b compares the contours of the switching region based on the ratio of the lattice signals, and the HCDW/CCDW peaks, revealing that the two are also spatially correlated.
Thus, also the lattice peak contraction can be used as a fingerprint to locate the HCDW switching. 

\section{Finite element method simulations}
Numerical simulations were performed using the COMSOL Multiphysics software, which uses the finite element method to solve partial differential equations governing heat conduction in solid materials and devices: 
\begin{align}
\rho C_{\rm p} \frac{\partial T}{\partial t}-\nabla \cdot(k \nabla T)=Q_{\rm e},
\end{align}
\begin{align}
Q_{\rm e}=\sigma|\nabla V|^2,
\end{align}
where $T$ is the temperature, $V$ the volume and $Q_{\rm e}$ the heat source term. The density of the material $\rho$, heat capacity $C_{\rm p}$ \cite{Suzuki1985}, as well as the $k$ thermal \cite{Nunez1985} and $\sigma$ electrical conductivity of all materials in the device were taken from literature or experiments \cite{Stojchevska,Mihailovic2021}.
The thermal model was coupled to the electromagnetic one which provided $Q_{\rm e}$ in the differential equations by calculating the resistive losses in the device, \textit{i.e.} Joule heating. 

\begin{figure*}[htpb]
  \centering
    \includegraphics[width=\linewidth, trim=0cm 0cm 0cm 0cm, clip=true]{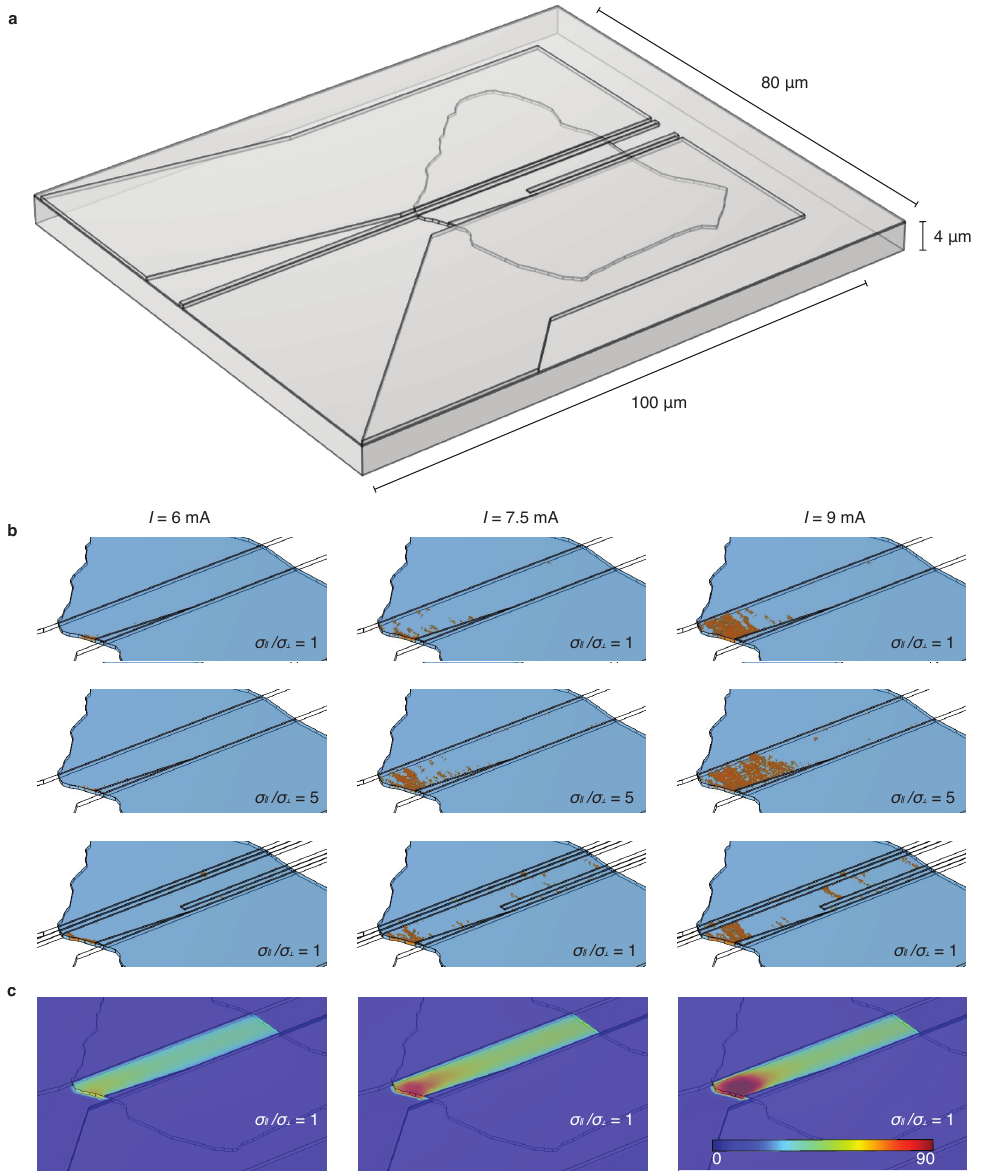}
    \caption{\textbf{Finite element method simulations}. \textbf{a} Schematic of the device considered for the simulation. \textbf{b} Switching volume (orange) under different current excitation levels (columns), as well as  assuming different material isotropy (top and middle row) and inclusion of a more detailed electrode geometry (bottom row). \textbf{c} Device heating upon switching for top row in b (color scale in Kelvin). 
    }
    \label{fig: S9}
\end{figure*}

The structure and dimensions of the simulated device is shown in Fig.~\ref{fig: S9}a. The geometry of the Au contacts and the 1$T$-TaS$_2$ crystal flake was realistically replicated both in shape and size, including irregularities such as the narrower spacing at one (left) side of the structure and the two additional disconnected contact fingers in the middle. Excitation of the device was applied via the two outer contacts by assuming \SI{100}{\micro \second} square wave current pulses of amplitudes between \si{0} and \SI{10}{\milli \ampere}.
At the beginning of each simulation, the device was fully in the unswitched state \textbf{A} at \SI{4}{\kelvin} temperature. After applying the excitation, switching of the elements from \textbf{A} to the fully-switched state \textbf{C} was handled by a simplified model which predicts formation of the HCDW state domain wall network under the influence of a sufficiently large charge carrier injection. Thus, in the present simulations the state of each element was switched once the current density of the element reached a predefined current density threshold of \SI{5e8}{\ampere / \meter\squared}, calibrated by the experiment.
Figure~\ref{fig: S9}b shows the switched material (orange) for different magnitudes of current excitation. All cases identify the preferential area of device switching at the left side of the device, which is influenced primarily by the narrower spacing between the contacts. With increasing current amplitude, the area gradually expands along the intergap space. Furthermore, we observe that the penetration depth of the switching is determined by the level of anisotropy of 1$T$-TaS$_2$. In the isotropic case (Fig.~\ref{fig: S9}b, top row), the switched volume propagates throughout the entire thickness of the crystal flake, whereas it remains closer to the surface when assuming an out/in-plane conductivity anisotropy of 1:5 (Fig.~\ref{fig: S9}b, middle row). A final set of simulations (Fig.~\ref{fig: S9}b, bottom row), we tested also the influence of the disconnected contact fingers. While they do affect the quantitative extent of the switched volume, the additional conductive paths they provide qualitative agree with the simulations in the top row of Fig.~\ref{fig: S9}b.

Contrary to the experiment these simulations do not reproduce the widening of the switched volume laterally and beneath the contacts. We speculate that this effect in the experiment is driven by propagation of local strain between the switched and unswitched area---a~feature presently not included in the numerical simulations. We also observe that compared to the experiment, the device switches fully at a lower current amplitude (between \si{10} to \SI{20}{\milli \ampere}). We attribute this to slight discrepancies between the assumed and effective material parameters.
Finally, in Fig.~\ref{fig: S9}c, device heating due to the current excitation is also demonstrated for the isotropic situation without influence of the disconnected middle contacts (Fig.~\ref{fig: S9}b, top row). The uneven heating upon current pulse application localized to the left side of the device is also in agreement with the preferential path of the current flow across the 1$T$-TaS$_2$ flake. 



\end{document}